\documentclass[]{aastex631}

\begin{document}

\title{GJ\,9404\,b: a confirmed eccentric planet, and not a candidate}

\author[0000-0002-3300-3449]{Thomas A. Baycroft}
\affiliation{School of Physics \& Astronomy, University of Birmingham, Edgbaston, Birmingham B15 2TT, United Kingdom}

\author{Harry Badnell}
\affiliation{School of Physics \& Astronomy, University of Birmingham, Edgbaston, Birmingham B15 2TT, United Kingdom}

\author{Samuel Blacker}
\affiliation{School of Physics \& Astronomy, University of Birmingham, Edgbaston, Birmingham B15 2TT, United Kingdom}

\author[0000-0002-5510-8751]{Amaury H.M.J. Triaud}
\affiliation{School of Physics \& Astronomy, University of Birmingham, Edgbaston, Birmingham B15 2TT, United Kingdom}



\begin{abstract}
Eccentric orbits can be decomposed into a series of sine curves which affects how the false alarm probability is computed when using traditional periodograms on radial-velocity data. Here we show that a candidate exoplanet orbiting the M dwarf GJ\,9404, identified by the HADES survey using data from the HARPS-N spectrograph, is in fact a bona-fide planet on a highly eccentric orbit. Far from a candidate, GJ\,9404\,b is detected with a high confidence. We reach our conclusion using two methods that assume Keplerian functions rather than sines to compute a detection probability, a Bayes Factor, and the FIP periodogram. We compute these using nested sampling with {\tt kima}.

\end{abstract}



\section{Introduction} \label{sec:intro}
Recently, \cite{2022HADES} reported partial results of the HADES survey, where they analyse the radial-velocity monitoring of 56 nearby M dwarfs stars obtained using the HARPS-N spectrograph at TNG (La Palma). The stars range from spectral types M0 to M3, and mass $0.3 ~{\rm M_{\odot}} < M_{*} < 0.71 ~\rm M_{\odot}$. 
Radial-velocities were extracted using the {\tt TERRA} template matching algorithm \citep{Anglada-Escudé_2012}. Planets are identified if they exceed a standard False Alarm Probability (FAP) of 0.1\%, calculated from a generalized Lomb-Scargle periodogram. \citet{2022HADES} list 11 planets detected by the HADES survey in several publications.

In order to compute occurrence rates, \citet{2022HADES} additionally identify  five more systems which they call `candidates' mentioning they need additional, more in-depth analysis. To find these, they perform a periodogram analysis along with, in some cases, a Gaussian process (GP) regression in order to model the effects of stellar rotational periods, which are prominent in M dwarfs. Stellar noise correction from activity indexes \citep[as in][]{Silva2011} are also used to remove observations affected by flares, and to detrend radial-velocity time series from magnetic cycles. 
One of these candidates is GJ\,9404\,b, with an orbital period $P = 13.46^{+0.01}_{-0.51}~\rm day$  and mass of $m_{\rm p}sin(i) = 10.3^{+1.8}_{-1.8}~\rm M_\oplus$. The star has a mass of \(M_{\star}=0.62\pm0.07\mathrm{\,M_{\odot}}\). Calculating a FAP requires a periodogram, which is a Fourier decomposition. As such \citet{2022HADES} assume sinusoidal (circular) radial-velocity modulations to assess a planet's detection, however GJ\,9404\,b is eccentric. 

In order to test a new method to compute detection limits \citep[as in][]{Standing2022} to be used to estimate occurrence rates, we reanalysed the entire HADES sample with the {\tt kima} nested sampling algorithm \citep{faria_kima}. As part of that re-analysis we find that GJ\,9404\,b is not just a candidate, but can be confirmed as a planet.

\section{Kima analysis \& results} \label{sec:methods}

For the radial velocity analysis we use the package {\tt kima} \citep{faria_kima}. The model includes Keplerian orbits for planets, a systemic velocity, and a jitter added in quadrature to account for any unmodelled variability. A Student's t-distribution is used for the likelihood evaluations. Compared to a Gaussian distribution, this naturally accounts for any outliers present in the data. The {\tt kima} package uses diffusive nested sampling \citep{brewer_dnest4} to explore parameter space and the number of Keplerian orbits is a free parameter. This allows for a Bayesian model comparison between models with different numbers of planets using a Bayes Factor. We take a Bayes Factor, \(\rm BF >150\) as very strong evidence in favour of the more complex model \citep[as in][]{kass_bayes_1995}. We also use the samples obtained by the nested sampling to compute a False-Inclusion Probability (FIP) periodogram \citep[as described in][]{Hara_FIP}. 
We take a FIP \(<0.01\) as a detection threshold. Given a series of detections with FIP \(<0.01\) we would expect \(<1\%\) of them to be false detections \citep{Hara_FIP}. At first sight this threshold might appear more permitting than the FAP, but FAPs are sensitive to false positives, whereas the FIP is a more reliable metric \citep{Hara_FIP}.

\begin{figure}
    \centering
    \includegraphics[width=.4\columnwidth]{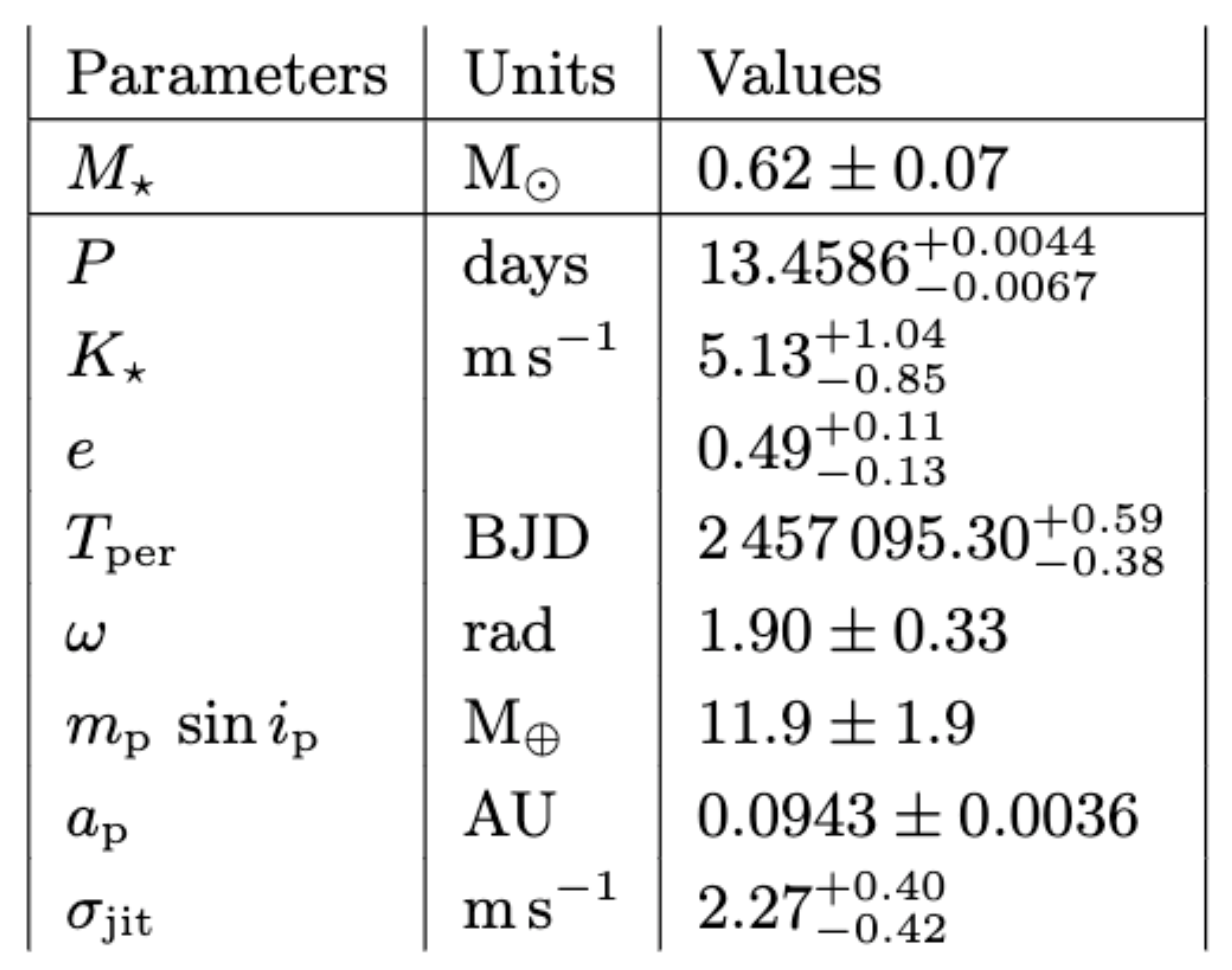}\\
    \includegraphics[width=.49\columnwidth]{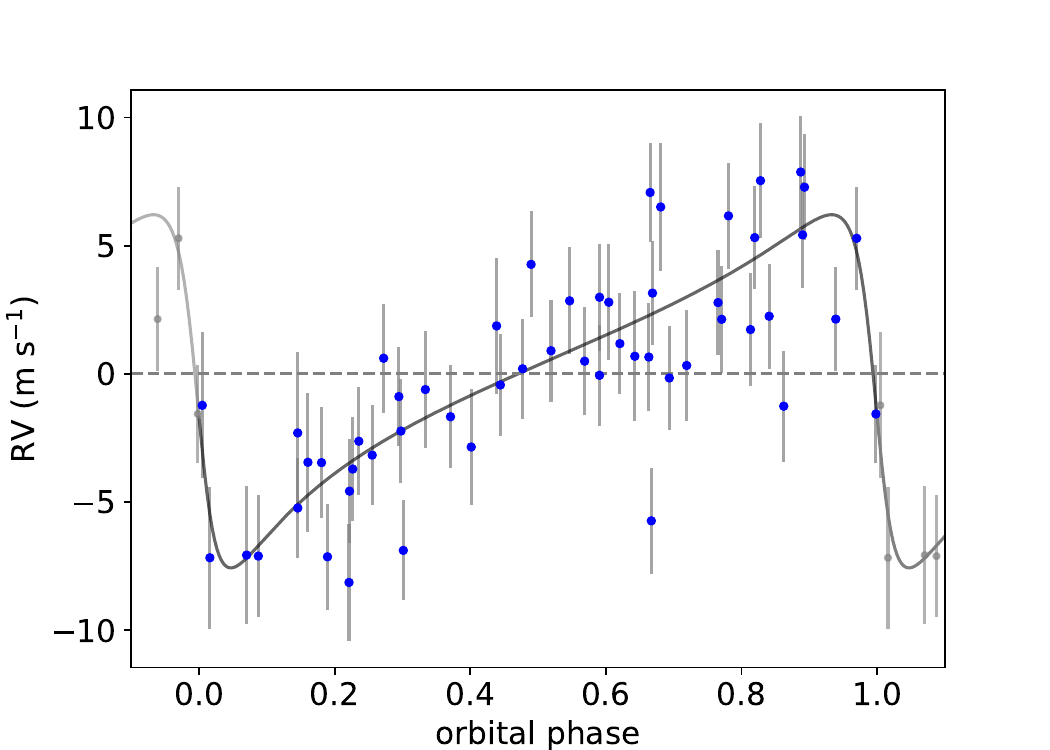}
    \includegraphics[width=.49\columnwidth]{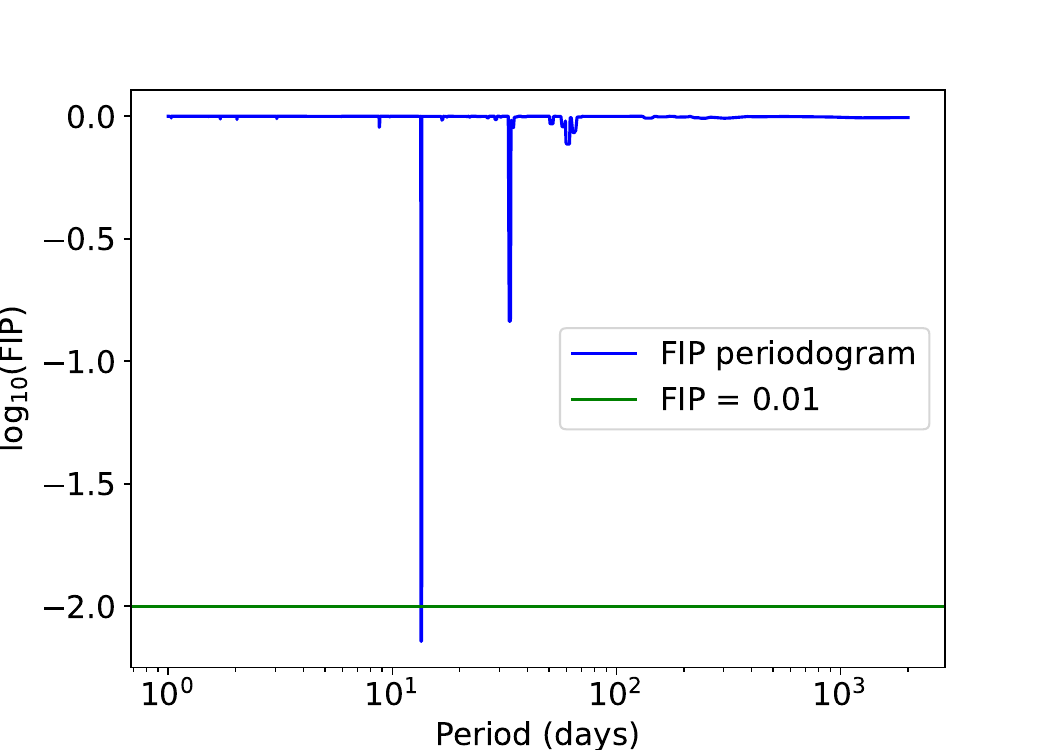}
    \caption{Top: Median values and \(1\sigma\) uncertainties for the parameters of GJ9404b from {\tt kima}. Bottom left: phase plot of best-fitting planetary solution. Bottom right: in blue the false-inclusion probability (FIP) periodogram, in green the threshold at which the false-incusion probability is 0.01.}
    \label{fig:plots}
\end{figure}

We reanalyse with {\tt kima} the radial velocity data for GJ\,9404 which we obtained from \citet{2022HADES}. This analysis results in $\rm BF = 782$ for a 1-planet model vs a 0-planet model, and $\rm BF = 7$ in favour of a 2-planet model vs a 1-planet model. This means that there is conclusive evidence for one Keplerian signal and moderate evidence for a second. The best-fitting solution is shown on the bottom left of Figure \ref{fig:plots}. In contrast to the analysis used in the \citet{2022HADES}, the use of Keplerian orbits within {\tt kima} allows for eccentric orbits.

The false-inclusion probability periodogram is shown on the bottom right of Figure \ref{fig:plots}. This shows two periodicities, one of which passes the detection threshold. The more prominent signal in the FIP periodogram corresponds to the best-fitting solution and the posterior distributions for the parameters are shown in the table in Figure \ref{fig:plots}. We note that this planet was claimed as a candidate in \citet{2022HADES} and that our period and mass values are compatible to the parameters presented in that publication. The rotation period of the star is $23.2 \pm 0.1$ days \citep{2022HADES}, so the signal we confirm is likely not stellar activity, and thus of planetary origin.

\section{Conclusion} \label{sec:results}

The example of GJ\,9404 highlights the limitations of traditional periodograms to identify exoplanetary signals in radial-velocity data and measure the robustness of their detections. Our re-analysis of the HADES data \citep{2022HADES} using {\tt kima} \citep{faria_kima, Baycroft_kima} reveals that GJ\,9404 hosts at least one exoplanet with a very high degree of confidence from the Bayes Factor. We confirm this by performing a FIP periodogram \citep{Hara_FIP}, which assume Keplerian functions rather than sines. Fig.~\ref{fig:plots} shows the FIP periodogram confirming the Bayes Factor estimated by {\tt kima}. This research note should be seen a reminder of the limitation of traditional periodograms to assess detection confidence, and a  reminder that more powerful and accurate statistical tools now exist.


\begin{acknowledgments}
This research received funding from the European Research Council (ERC) under the European Union's Horizon 2020 research and innovation programme (grant agreement n$^\circ$ 803193/BEBOP).

\end{acknowledgments}

%

\vspace{5mm}
\facilities{TNG(HARPS-N)}


\software{  
          kima \citep{faria_kima}, DNEST4 \citep{brewer_dnest4}
          }










\bibliography{sample631}{}
\bibliographystyle{aasjournal}



\end{document}